\documentclass[10pt, conference]{IEEEtran}
\IEEEoverridecommandlockouts
\usepackage[space]{cite}
\usepackage{amsmath,amssymb,amsfonts}
\usepackage{textcomp}
\usepackage{listings}
\usepackage{xcolor}
\usepackage{graphicx} % Required for inserting images
\usepackage{algorithm}
\usepackage{algpseudocode}
\usepackage[labelfont=bf,figurename=Figure]{caption}
\usepackage{hyperref}

\usepackage{stfloats}
\usepackage{array}
\usepackage{listings}

\usepackage{booktabs} %nice tables
\usepackage{subfiles} % Best loaded last in the preamble

\newcommand{\ra}[1]{\renewcommand{\arraystretch}{#1}}
\newcommand{\secref}[1]{\S \textcolor{green}{\ref{#1}}}
\newcommand{\figref}[1]{\textcolor{green}{\ref{#1}}}
\newcommand{\tableref}[1]{\textcolor{green}{\ref{#1}}}

\definecolor{codegreen}{rgb}{0,0.6,0}
\definecolor{codegray}{rgb}{0.5,0.5,0.5}
\definecolor{codepurple}{rgb}{0.58,0,0.82}
\definecolor{backcolour}{rgb}{0.95,0.95,0.92}
\definecolor{darkred}{rgb}{0.55, 0.0, 0.0}

\hypersetup{
  colorlinks   = true, %Colours links instead of ugly boxes
  urlcolor    = darkred, %Colour for external hyperlinks
  linkcolor   = green, %Colour of internal links
  citecolor   = red %Colour of citations
}

\lstdefinelanguage{myc}{
  language     = C,
  morekeywords = {Queue, node, emucxl_alloc, emucxl_free, kv_store, kvs_obj, kvs_pair, NULL,
  POLICY1, POLICY2, LOCAL_MEMORY, REMOTE_MEMORY},
  float=tp,
  floatplacement=tbp,
  aboveskip=0pt,
  belowskip=0pt,
  lineskip=0pt
}

\definecolor{cornflowerblue}{rgb}{0.39, 0.58, 0.93}
\definecolor{darkbyzantium}{rgb}{0.36, 0.22, 0.33}
\definecolor{iris}{rgb}{0.35, 0.31, 0.81}
\definecolor{midnightblue}{rgb}{0.1, 0.1, 0.44}
\definecolor{royalblue(traditional)}{rgb}{0.0, 0.14, 0.4}
\definecolor{royalblue(web)}{rgb}{0.25, 0.41, 0.88}

\lstdefinestyle{mystyle}{
    backgroundcolor=\color{white},   
    commentstyle=\rmfamily\itshape\color{gray},
    keywordstyle=\bfseries\color{darkred},
    numberstyle=\tiny\color{codegray},
    stringstyle=\color{codepurple},
    stringstyle=\rmfamily,
    basicstyle=\ttfamily\footnotesize,
    breakatwhitespace=false,         
    breaklines=true,                 
    captionpos=b,                    
    keepspaces=true,
    frame=single,
    showspaces=false,                
    showstringspaces=false,
    showtabs=false,                  
    tabsize=2
}

\lstdefinestyle{customc}{
  belowcaptionskip=1\baselineskip,
  breaklines=true,
  frame=L,
  xleftmargin=\parindent,
  language=C,
  showstringspaces=false,
  basicstyle=\footnotesize\ttfamily,
  keywordstyle=\bfseries\color{green!40!black},
  commentstyle=\itshape\color{purple!40!black},
  identifierstyle=\color{blue},
  stringstyle=\color{orange},
}

\def\emucxl{\textrm{emucxl}}
\def\Emucxl{\textrm{emucxl }}

\IEEEoverridecommandlockouts

\begin{document}
% standardized
% \title{\emucxl: an emulation framework for CXL-based 
% disaggregated memory applications}
% % \title{\emucxl: an access library for disaggraged memory using
% % emulated CXL}
% \author{\IEEEauthorblockN{Raja Gond}
% \IEEEauthorblockA{\textit{Department of Computer Science and Engineering} \\
% \textit{Indian Institute of Technology Bombay}\\
% Mumbai, India \\
% rajagond@cse.iitb.ac.in
% }
% \and
% \IEEEauthorblockN{Prof. Purushottam Kulkarni}
% \IEEEauthorblockA{\textit{Department of Computer Science and Engineering} \\
% \textit{Indian Institute of Technology Bombay}\\
% Mumbai, India \\
% puru@cse.iitb.ac.in
% }
% }

% \maketitle
% % \IEEEpeerreviewmaketitle
% \thispagestyle{empty}
% \setcounter{page}{1}

% \bstctlcite{IEEEDAGit:BSTcontrol}

\title{\emucxl: an emulation framework for CXL-based disaggregated memory applications}
\author{
\IEEEauthorblockN{Raja Gond and Purushottam Kulkarni}
\IEEEauthorblockA{\textit{Department of Computer Science and Engineering}\\
\textit{Indian Institute of Technology Bombay}\\
\{rajagond, puru\}@cse.iitb.ac.in}
}
\maketitle
% \IEEEpeerreviewmaketitle
\thispagestyle{empty}
\setcounter{page}{1}
%%%%%%%%%%%%%%%%%%%%%%%%%%%%%%%%%%%%%%%%%%%%%%%%%%%%%%%%%%%%%%%%%
%%%%%%%%%%%%%%%%%%%%% ABSTRACT %%%%%%%%%%%%%%%%%%%%%%%%%%%%%%%%%%
%%%%%%%%%%%%%%%%%%%%%%%%%%%%%%%%%%%%%%%%%%%%%%%%%%%%%%%%%%%%%%%%%
\begin{abstract}

The emergence of CXL (Compute Express Link) promises to transform the status of interconnects between host and devices and in turn impact the design of all software layers. With its low overhead, low latency, and memory coherency capabilities, CXL has the potential to improve the performance of existing devices while making viable new operational use cases (e.g., disaggregated memory pools, cache coherent memory across devices etc.). The focus of this work is design of applications and middleware with use of CXL for supporting disaggregated memory. A vital building block for solutions in this space is the availability of a standard CXL hardware and software platform. Currently, CXL devices are not commercially available, and researchers often rely on custom-built hardware or emulation techniques and/or use customized software interfaces and abstractions. These techniques do not provide a standard usage model and abstraction layer for CXL usage, and developers and researchers have to reinvent the CXL setup to design and test their solutions, our work aims to provide a standardized view of the CXL emulation platform and the software interfaces and abstractions for disaggregated memory. This standardization
is designed and implemented as a user space library, \Emucxl and is available as a virtual appliance. The library provides a user space API and is coupled with a NUMA-based CXL emulation backend. Further, we demonstrate usage of the standardized API for different use cases relying on disaggregated memory and show that generalized functionality can be built using the open source \Emucxl library.

\end{abstract}


%%%%%%%%%%%%%%%%%%%%%%%%%%%%%%%%%%%%%%%%%%%%%%%%%%%%%%%%%%%%%%%%%%%%%%%%%%%%%%
%%%%%%%%%%%%%%%%%%%%%%%%%%%%%%%% INTRODUCTION %%%%%%%%%%%%%%%%%%%%%%%%%%%%%%%%%%
%%%%%%%%%%%%%%%%%%%%%%%%%%%%%%%%%%%%%%%%%%%%%%%%%%%%%%%%%%%%%%%%%%%%%%%%%%%%%%
% \begin{itemize}
%     \item Black - Okay!!
%     \item \textcolor{iris}{Iris} - Change required
%     \item \textcolor{teal}{Teal} - Draft (Review Required)
% \end{itemize}
\section{Introduction}
\label{section:introduction}
% \subfile{sections/introduction}
Data is gold and utility of data lies 
in a data-centric approach to computing --- big 
data~\cite{bigdata}, machine learning ~\cite{aiml}, stream 
processing~\cite{aiml}, data stores and transaction systems 
~\cite{in-memory}, to name a few. A vital component of these 
modern systems is the rate at which data can be accessed and 
stored, which in turn is dependent on the interconnect technology
to move data between compute, memory and storage devices.

Typical data center capacities revolve around large storage 
resources and clusters of compute nodes that access this storage
via high speed network connections. While techniques like
data-hoarding~\cite{hoarding-prefetching}, data-prefetching~\cite{data-prefetching, hoarding-prefetching}, caching~\cite{caching}
write back-based updates~\cite{write-back}  provide 
in-memory performance benefits for applications, but they are limited by the 
disparity in the available memory on a machine and the size of the 
accessed data store. Additionally, maintaining two copies of the data
requires software-based cache coherency mechanisms, which can offer
different trade-offs of performance vs. cost (of maintaining coherency
or of data loss). 
A popular trajectory of work has explored the design of in-memory 
(or disaggregated) distributed systems~\cite{ramcloud, distributedKV} 
% with high speed interconnects (~\cite{rdma, rdma2,
% rdma, infiniband, gigabitethernet, smartnics}).
with high speed interconnects~\cite{rdma, rdma2, infiniband, fast-ethernet, azure-smartnic}.
While these systems overcome the capacity requirements to quite some 
extent, they still rely on software based mechanisms for caching and 
coherency mechanisms. 

Of late, Compute Express Link (CXL)~\cite{cxl} is garnering 
attention from both industry and academia. CXL is an open industry-supported standard designed as a hardware-assisted cache-coherent interconnect for processors, memory, and accelerators with PCIe as
the physical layer interconnect. 
The CXL standard specifies different interconnect modes, one of 
which enables memory expansion for disaggregated memory 
solutions~\cite{dd}, %~\cite{cxl.mem}
an operating mode to access disaggregated memory with NUMA-level
access latencies~\cite{pond,Maruf_2023}. %~\cite{numamcxl}
An important challenge that arises with the availability of new 
hardware support is facilitating the smooth integration of software changes.
Several research and industry efforts 
% are already underway~\cite{a,b,c,d, 
are already underway~\cite{linuxcxl, qemucxl, libcxl, smdk, smt, Maruf_2023}
to best utilize and design CXL-centric software (applications, middleware,
OS-updates) to best utilize CXL's feature set. 

An important missing piece (at least currently) in the above story
is the non-availability of CXL hardware. While retail CXL hardware
is projected to be available soon~\cite{samsungcxl2.0}, the development
of standards, tools and CXL-based solutions has already found momentum. Two main challenges these efforts face is that 
a non-standardized access interface and abstractions for CXL memory
and a non-standardized simulation and emulation setup. 
Typically, each CXL-based development effort has developed
custom interfaces for accessing CXL functionality and further
developed custom simulation, emulation or custom hardware 
prototypes for testing and demonstration. 
While these serve the purpose for each solution, the lack of 
standardization and open-source availability leads to redundant 
efforts for each work to independently setup its platforms
and a non-uniform domain for comparison across solutions.

As part of this work, we design and build \emucxl, an emulation
framework for CXL-based disaggregated memory solutions. Our solution,
provides a configurable CXL emulation setup as a virtual appliance,
i.e., as a virtual machine with a configurable setup that 
emulated CXL behaviour via customized configuration of virtual 
and physical NUMA resources. The virtual appliance setup is based
on
% \raja{Removing this texts: and extends}  
the ideas proposed by POND~\cite{pond}.
Further, \Emucxl provides an API in user space to access and
utilized CXL based disaggregated memory. The implementation of the 
API sits along with OS-level memory management and uses the 
NUMA layout for utilizing the disaggregated emulation setup
of CXL memory. We envision developers and researchers to use 
the \Emucxl virtual appliance out-of-the-box as a CXL 
emulation setup and the \Emucxl library for rapid prototyping
of user-space solutions. 
We demonstrate
the usage of \Emucxl via two types of uses cases---an application
using the raw \Emucxl API and a middleware that used the \Emucxl
API to provide higher-level abstractions for applications to use.
The claim of \Emucxl is it simultaneously provides a standard API 
to build solutions and in term compare and contrast them
for features and performance in a more uniform manner and 
drastically reduces the CXL platform and access abstraction
setup for developers and researchers.
The contributions of our work towards these claims are 
as follows,
\begin{enumerate}
    \item[(i)] Making available a virtual appliance setup for emulation of
    CXL based disaggregated memory.
    \item[(ii)] Design and implementation of an user-space API for access
    and usage of CXL memory.
    \item[(iii)] Demonstration via two different types of use cases---direct
    application access and application middleware---the feature set
    and usage examples of the \Emucxl API.
    \item[(iv)] Empirically demonstrate the correctness and impact of 
    the emulated CXL disaggregated memory operations.
    \item[(v)] An open-source virtual appliance setup to aid rapid prototyping of
    CXL-based disaggregated memory solutions, and an API specification 
    and implementation that is extendable.
\end{enumerate}

The remainder of this paper is organized as follows, 
Background and motivation are discussed in 
Section~\secref{section:background_and_motivation}, the design and implementation are discussed in Section~\secref{section:design_and_implementation}, and example use cases of \Emucxl are discussed in 
Section~\secref{section:use_cases}. 
Related work is summarized in Section~\secref{section:related_works} 
and Section~\secref{section:conclusion} concludes the paper with 
a brief discussion of future work.

%%%%%%%%%%%%%%%%%%%%%%%%%%%%%%%%%%%%%%%%%%%%%%%%%%%%%%%%%%%%%%%%%%%%%%%%%%%%%%
%%%%%%%%%%%%%%%%%%%%%%%%%%%%%% Background and Motivation %%%%%%%%%%%%%%%%%%%%%
%%%%%%%%%%%%%%%%%%%%%%%%%%%%%%%%%%%%%%%%%%%%%%%%%%%%%%%%%%%%%%%%%%%%%%%%%%%%%%

\section{Background \textit{\&} Motivation}
\label{section:background_and_motivation}
%\subfile{sections/background}
As the scale of compute and data storage increases, scaling DRAM 
technology is becoming increasingly challenging from performance, utilization and cost perspectives~\cite{dram-challenge, dram-challenge2, pmem-dram-challenge}.
One of the many solutions being actively explored 
is \textit{disaggregated memory}~\cite{dmem, mem.dis}--memory that is provisioned from a pool of memory hardware. The main idea is that a shared pool can provide
flexible and increased provisioning capacity while also improving utilization.
The main challenge with this approach is to provide remote memory
access with low latency and high throughput. In this regard, the device interconnect technology plays a critical role. Several interconnect technologies 
are on offer --- PCIe~\cite{pcie}, Infiniband~\cite{infiniband2, infiniband}, RoCE~\cite{roce}, Fast Ethernet~\cite{fast-ethernet}, RDMA~\cite{rdma}, NVLink~\cite{nvlink}, NVSwitch~\cite{nvswitch} etc.
Each one provides different abstractions for operations, e.g., PCIe based
access is byte or block addressable, the RDMA and Ethernet based options operate on the IO path, and NVLink provides a fast interconnect between GPUs and/or CPUs. Providing a disaggregated memory abstraction on top of these technologies
implies building an additional layer for converting memory to IO 
or device access, and handling data coherency semantics across data instances
~\cite{infiniswap, rdma2, directcxl}. %reference taken from directcxl paper

%%%%%%%%%%%%%%%%% 2nd Paragraph %%%%%%%%%%%%%%%%%%%%%%%%%%%%%%
\begin{table}[tp]
\centering
\ra{1.3}
\begin{tabular}{@{}p{4cm}p{1.5cm}p{1.2cm}p{1cm}@{}}
\toprule
\textbf{Solution} & \textbf{\textit{Implementation}}& \textbf{\textit{Availability} } & \textbf{\textit{Standard API}}\\
\midrule
Two NUMA nodes---one with memory and CPU, the other cpu-less and only memory.
\cite{pond} &
Software (emulation) & \href{https://github.com/vtess/Pond}{Open source} & No \\
\hline

Customized memory add-in-cards, 16nm FPGA-based processor nodes, 
CXL switch and controller, and a PCIe backplane. 
\cite{directcxl} &
Hardware prototype & Custom & Yes \\
\hline
CXL.mem and CXL.io via 16nm FPGA boards integrated with in-house RISC-V CPU 
and NVMe storage, connected via a tailored PCIe backplane.  
\cite{pcie} & Hardware prototype & Custom & No \\
\hline
Customized FPGA boards with CXL PHY link to CPU, CXL controller
for CXL.mem and CXL.io and a DDR3 DIMM module.
\cite{in-memory} & Hardware prototype & Custom & No \\
\hline
CXL-based memory and storage\cite{deeplearning}  via NUMA nodes and creation 
of RAMDisks~\cite{ramdisk} on remote nodes.
& Software (emulation) & Custom & No \\
\hline
CXL-enabled memory pooling using NUMA domains and 
integration with the \textrm{libnuma} library~\cite{wahlgren2022}.
& Software (emulation) & \href{https://github.com/KTH-ScaLab/mem-emu}{Open source} & No \\

\bottomrule

\end{tabular}
\caption{Characteristics of Hardware and Software Based CXL platforms.}
\label{table:cxl-platforms}
\end{table}

Compute Express Link (CXL)~\cite{cxl}, an emerging interconnect technology,
promises to provide multiple abstractions as access semantics 
and hardware-assisted cache coherency for distributed data accesses.
The hardware-assisted cache coherency support has the potential to 
greatly simplify the software stack for data coherency handling 
and enable high performance distributed data applications.

\begin{figure}
    \centering
    \includegraphics[width=0.45\textwidth]{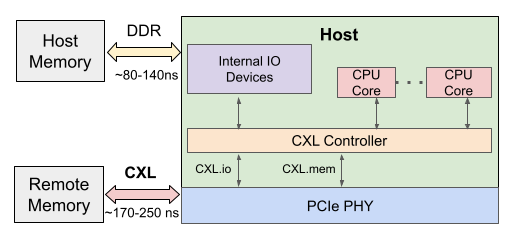}
        \caption{Architecture and components of CXL-based disaggregated memory.}
    \label{fig:cxl-archi}
\end{figure}
Figure~\figref{fig:cxl-archi} shows the architecture of a 
typical CXL-enabled system using disaggregated memory. 
The host and remote memory together constitute the 
total physical memory, and the CXL controller executes
the CXL protocols, multiplexing and management of the 
memory resources. All CPU issued load and store instructions 
to access  remote memory pass 
through a PCIe bus
via the CXL controller~\cite{mem.dis}.
CXL defines three protocols to provide different functionalities~\cite{cxl1.0, dd}. 
\begin{itemize}
    \item \textbf{CXL.io}: A mandatory protocol to set up 
device connections with the host, with primitives for device discovery, 
device configuration and reconfiguration.
    \item \textbf{CXL.cache}: An optional protocol that enables a
device to access and cache data from the host attached memory,
and maintain coherency between the host memory and memory on 
attached devices.

    \item \textbf{CXL.mem}: A protocol to be used by a host machine
to map device memory on to its own address space. 
Doing so, the host machine can issue direct memory load/store 
instructions to access and modify remote device memory while 
ensuring memory coherency. 
Integrating load/store instructions with CXL.mem also implies
that the CPU can cache contents of the PCIe attached 
memory in at level of its caches.

\end{itemize} 
The CXL controller via its transport protocol arbitrates link access between
the three protocols and 
achieves transfer rates up to 32 GB/s and 64 GB/s in each 
direction over a 16-lane link, for PCIe5.0 and PCIe6.0
physical layers, respectively~\cite{dd}.

The scope of this work is to develop an emulation and access 
framework for  utilizing functionality provided via the CXL.io 
and CXL.mem protocols.

Following the release of the CXL specification, several researchers
and developers have responded by studying implications on performance
and design of solutions for usage with applications. Due to the absence 
of commercially available CXL hardware, each of these solutions has 
resulted in the development of custom CXL access interfaces and emulation platforms.
Table \ref{table:cxl-platforms} summarizes existing solutions and their features. 
As can be seen, all the solutions either develop and use custom hardware prototypes (FPGA-based) or use software based
emulation/simulation for CXL functionality. Further, each of them uses
a non-standard access API to interface and consume the CXL functionality.

Our claim is that the lack of standardization makes it challenging to establish 
an unified approach for exploring, extending and comparing CXL implications
and CXL-based applications, middleware and the operating system solutions.

The focus of this work is to address these challenges and specifically
those related to the software interfaces and abstractions, and emulation
platform for disaggregated memory usage.
We aim to provide a framework via the \Emucxl library 
for a standardized access layer and a virtual appliance setup for
a out-of-the-box CXL emulation platform.

The aim of \emucxl, our proposed emulation framework, is to enhance 
interoperability of solutions, simplify development efforts, and 
enable researchers and developers to explore, extend and evaluate 
behavior of CXL-enabled systems efficiently.
Without the need to build or reinvent custom CXL platforms and access interfaces,
a significant amount of time, effort, and resources, can be
saved and utilized for building CXL-based solutions.

Towards providing a standardized CXL emulation framework
for CXL based disaggregated memory applications, we address
the following requirements,
\begin{itemize}
    \item The framework should consist of both an emulation layer
    for CXL disaggregated memory semantics and an user-level interface
    for leveraging CXL semantics of local and remote memory. 
    \item The user-level interface should support actions related
    to configuration of disaggregated memory---static and dynamic sizing 
    of local and remote memory, movement of data between disaggregated 
    memory nodes, reporting of memory usage/mappings, policies for 
    data movement across local and remote memory regions, and so on.
    \item Provide an integrated run time environment of the CXL emulation
    setup and the CXL access interface, so that developers can use 
    the interface and its implementation without worrying about any
    of the detailed CXL setup and emulation tasks.
\end{itemize}

%%%%%%%%%%%%%%%%%%%%%%%%%%%%%%%%%%%%%%%%%%%%%%%%%%%%%%%%%%%%%%%%%%%%%%%%%%%%%%
%%%%%%%%%%%%%%%%%%%%%%%%%%%%%%%% Scope of Work %%%%%%%%%%%%%%%%%%%%%%%%%%%%%%%%%%
%%%%%%%%%%%%%%%%%%%%%%%%%%%%%%%%%%%%%%%%%%%%%%%%%%%%%%%%%%%%%%%%%%%%%%%%%%%%%%

% \section{Problem Statements}
% \label{section:problem_statements}
% \section{Scope of work}
% \label{section:scope}

% \subfile{sections/scope}
% \input{sections/scope}

%%%%%%%%%%%%%%%%%%%%%%%%%%%%%%%%%%%%%%%%%%%%%%%%%%%%%%%%%%%%%%%%%%%%%%%%%%%%%%
%%%%%%%%%%%%%%%%%%%%%%%%%%%%%%%% Design and Implementation %%%%%%%%%%%%%%%%%%%
%%%%%%%%%%%%%%%%%%%%%%%%%%%%%%%%%%%%%%%%%%%%%%%%%%%%%%%%%%%%%%%%%%%%%%%%%%%%%%

\section{Design \textit{\&} Implementation}
\label{section:design_and_implementation}

\begin{figure}
    \centering
    \includegraphics[width=0.35\textwidth]{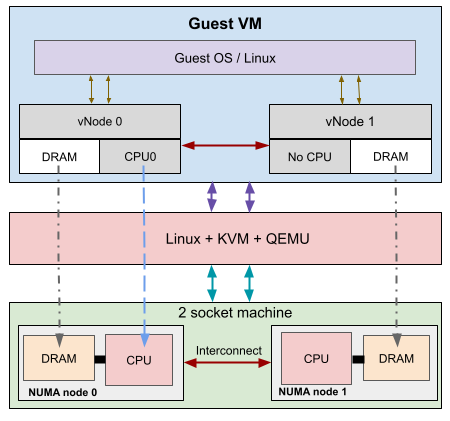}
    \caption{Resource components and mapping of NUMA based CXL emulation of disaggregated memory.}
    \label{fig:pondvm}
\end{figure}
In this section, we discuss the design and implementation 
details of the user space library, \emucxl\footnote{\Emucxl source code is publicly available at \url{https://github.com/cloudarxiv/emucxl}, which includes source files for the library implementation, setup instructions, and various use cases.}.
\Emucxl extends the CXL-emulation solution to provide
virtual appliance---a virtual machine with the emulation 
setup and a software setup with a standardized API 
for CXL-style memory pooling.
Furthermore, \Emucxl aims to benefit developers and researchers
by offering a solution that avoids the need to redevelop interfaces
and emulation platforms for CXL-based memory pooling solutions.
The setup of the virtual appliance is shown in Figure~\figref{fig:pondvm},
and uses ideas proposed by POND~\cite{pond} for the CXL 
disaggregated memory emulation.
The main requirements of features of the setup are as 
follows---
\begin{itemize}
    \item a 2-socket physical server as the underlying hardware.
    \item a two-node qemu+kvm virtual machine.
    \item one vNode of the VM mapped to a physical node 
    of the 2-socket physical machine.
    \item the second vNode of the VM mapped to physical 
    memory of the second socket of the physical node,
    while the vCPUs of the second node are not mapped,
    and in effect the second vNode is cpuless.
\end{itemize}
The CPU can access the disaggregated memory directly
via load/store semantics. The standardized API for 
disaggregated memory was designed with a set of guiding
principles in mind to ensure ease of use, portability, 
and flexibility. The API exposes a comprehensive range of functionality that allows for fine-grained management over memory resources. Memory allocation and deallocation, 
as well as data migration between local and remote memory,
are important features. Furthermore, the API design
incorporates extensibility to accommodate others/evolving
application requirements of disaggregated memory.

The standardized API is designed and implemented as an user space
library, \emucxl, with a kernel backend. This library offers a
comprehensive user space API and is complemented by a NUMA-based
CXL emulation backend discussed above. The NUMA-based backend 
leverages NUMA-aware memory allocation semantics and is 
implemented as a Linux kernel module. To successfully deploy
the \Emucxl library and its accompanying backend, it is recommended 
to use an Ubuntu version, preferably 22.04, with a kernel version 
equal to or greater than 5.15. This software configuration ensures
compatibility for the successful execution of the \Emucxl functionality.

The kernel backend of \Emucxl is loaded and unloaded as a Linux Loadable
kernel module(LKM). During loading of \Emucxl kernel backend(LKM), a device file is created and registered within the kernel; during unloading, it is unregistered. This kernel backend provides
\verb|mmap()| functions for device driver discussed below.
%%%%%%%%%%%%%% API %%%%%%%%%%%%%%%%%
\begin{table*}[tp]
\centering
\ra{1.0}
\begin{tabular}{@{}
    >{\raggedright\arraybackslash}p{8.8cm} 
    >{\raggedright\arraybackslash}p{8cm}%
  @{}}
\toprule
\textbf{Standardized API} & \textbf{\textit{Description}}\\
\midrule
\verb|void emucxl_init()|& 
open CXL device file, store file descriptor, initialized emulated memory sizing\\
\verb|void emucxl_exit()| & free all allocated memory and close the device file  \\
\hline
\verb|void* emucxl_alloc(size_t size, int node)|
& allocate memory either remotely or locally \\
\verb|| & node = 0 for local memory, and 1 for remote memory \\
\verb|| & returns virtual address to calling process \\
\verb|void emucxl_free(void* address, size_t size)| 
& free allocated memory block of the specified size \\
\hline
\verb|void* emucxl_resize(void* address, size_t size)| & allocate memory of new size on same node, copy, free earlier allocation, return address \\
\verb|void* emucxl_migrate(void* address, int node)| &
allocate memory on specified node, migrate all data to new address and return address\\
\hline
\verb|bool emucxl_is_local(void* address)|\newline
\verb|int emucxl_get_numa_node(void* address)|\newline
\verb|size_t emucxl_get_size(void* address)|\newline
\verb|size_t emucxl_stats(int node)|
& check if address is mapped from 
local or remote memory \newline
return the NUMA node associated with the given memory address\newline
return size of memory allocated of the specified memory address\newline
return size of total memory allocated on given NUMA node\\
\hline
\verb|bool emucxl_read(void* addr, int, void* buf, size_t)|\newline
\verb|bool emucxl_write(void* buf, int, void* addr, size_t)|\newline
\verb|void* emucxl_memset(void* addr, int value, size_t)|
& read/write specified number of bytes from/to the memory address 
and store in/from buffer \newline 
fill a block of memory with either 0 or -1\\
\hline
\verb|void* emucxl_memcpy(void*, const void*, size_t)| % second arg is const
\verb|void* emucxl_memmove(void*, const void*, size_t)| % second arg is const
& copy data from source to destination address \newline 
move data from source to destination address \newline
unlike \textrm{memcpy}, \textrm{memmove} ensures correct handling of overlapping memory regions when moving the data. \\
\bottomrule
\end{tabular}
\caption{Description of standardized APIs of the \Emucxl library.}
\label{table:apis}
\end{table*}

\begin{figure}[t]
    \centering
    \includegraphics[width=0.5\textwidth]{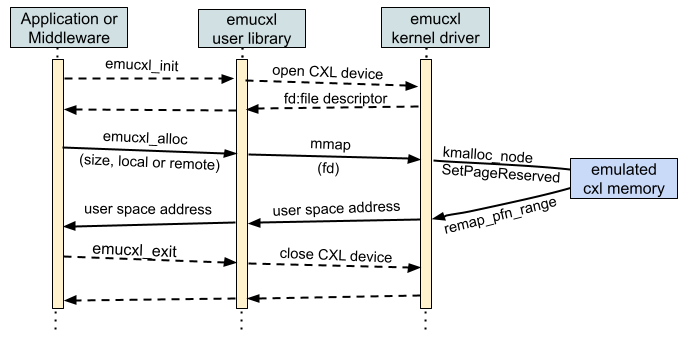}
 \caption{Message sequence diagram for initialization
and usage of the \Emucxl library, for allocation of disaggregated
memory. }
    \label{fig:msg_seq}
\end{figure}

Table~\tableref{table:apis} provides an overview of the standardized
APIs  we  defined and their brief descriptions. Most of these APIs are
implemented in user space, with few exceptions. Resource allocation is a 
critical aspect of the \Emucxl library. 
It provides \textit{emucxl\_alloc} and
\textit{emucxl\_free} APIs to allocate and 
deallocate memory on NUMA-based 
CXL emulation backend, respectively. 
Figure~\ref{fig:msg_seq} illustrates the sequence of calls
for \textit{emucxl\_init}(initialization),
\textit{emucxl\_alloc}(allocation of disaggregated memory) and
\textit{emucxl\_exit}(closes the device) APIs of \Emucxl library.

As shown, \textit{emucxl\_alloc} uses the \textit{mmap()}
system call on a file descriptor associated with the CXL device (Opened by the 
\textit{emucxl\_init} API call) to achieve the functionality. 
The device driver's \texttt{mmap()} operation has the following signature:\\
\texttt{int (*mmap)(struct file *filp, struct vm\_area\_struct *vma);}.\\
The \textit{filp} field is a pointer to a struct file created when the device is opened from user space. The \textit{vma} field is used to indicate the virtual address space where the memory should be mapped by the device~\cite{linux-memory-mapping}.
\\
Note that the traditional \verb|mmap| system call is not NUMA-aware, 
so we overrode \verb|mmap()| operation 
via device driver's file\_operations to address our specific needs for NUMA-aware allocation on NUMA-based
emulated CXL device. 
As shown in Figure~\ref{fig:msg_seq}, we use \textit{kmalloc\_node} Linux
kernel API to allocate memory on virtual NUMA Node, and
then map it to the 
user address space as indicated by the \textit{vma} parameter using helper 
functions such as \textit{remap\_pfn\_range()}. Since the pages are mapped to 
user space, they might be swapped out. To avoid this, we must set the 
\verb|PG_reserved| bit on the page, and it can be done using \textit{SetPageReserved()}
~\cite{linux-memory-mapping}.
\textit{kmalloc\_node} expects a NUMA node number (0 for local and 1 for remote) to allocate memory. There is no direct way to pass this information with
\texttt{mmap()} system call,\\ \texttt{void *mmap(void addr[.length], size\_t length, int prot, int flags, int fd, off\_t offset);}\\
(Note that the signature of this mmap system call differs from the signature of the earlier mmap device driver operation).
Therefore, we overloaded the
offset field to pass the node number from \textit{emucxl\_alloc} to the device driver
\verb|mmap| operation. \textit{emucxl\_free} uses the \verb|munmap()| system
calls to deallocate the memory. \textit{emucxl\_init} uses \verb|open| to
open the device file and \textit{emucxl\_exit} uses \verb|close| to
close the device file. All other API listed in Table~\tableref{table:apis} implementations are fairly straightforward and implemented in user space 
with basic \textbf{C} library. Metadata (i.e. address, size, NUMA node) of
each allocation/deallocation of \Emucxl library is maintained in the data structure which utilizes by \textit{emucxl\_is\_local, emucxl\_get\_numa\_node, emucxl\_get\_size} 
and \textit{emucxl\_stats} APIs for their implementation. The implementation of other APIs is also similar to what we do with \texttt{malloc} and
\texttt{free} in C, except here we are using \textit{emucxl\_alloc} 
and \textit{emucxl\_free}.

%%%%%%%%%%%%%%%%%%%%%%%%%%%%%%%%%%%%%%%%%%%%%%%%%%%%%%%%%%%%%%%%%%%%%%%%%%%%%%
%%%%%%%%%%%%%%%%%%%%%%%%%%%%%%%% USE CASES %%%%%%%%%%%%%%%%%%%%%%%%%%%%%%%%%%
%%%%%%%%%%%%%%%%%%%%%%%%%%%%%%%%%%%%%%%%%%%%%%%%%%%%%%%%%%%%%%%%%%%%%%%%%%%%%%

\section{\Emucxl framework use cases}
\label{section:use_cases}

\Emucxl provides two main ways to use its standardized library 
for disaggregated memory based applications. The raw \Emucxl API
can be directly used by applications and manage memory from 
within the application. We label this usage as \textit{direct
access}, where the logic of managing local and remote memory
is part of the applications logic. 
An alternative mechanism for applications to use disaggregated 
memory is via interfaces provided by a middleware platform, which, in turn, uses the \Emucxl API and additional metadata
to support disaggregated memory based abstractions to applications.
The memory management optimizations of handling local and remote memory efficiently are responsibility of the middleware. We label this second type of usage as \textit{middleware driven}
usage of \Emucxl.
As part of this work, we demonstrate two examples of middleware driven usage of \Emucxl---a key-value data store
and a slab allocator.

\begin{figure}[t]
    \centering
    \includegraphics[width=0.43\textwidth]{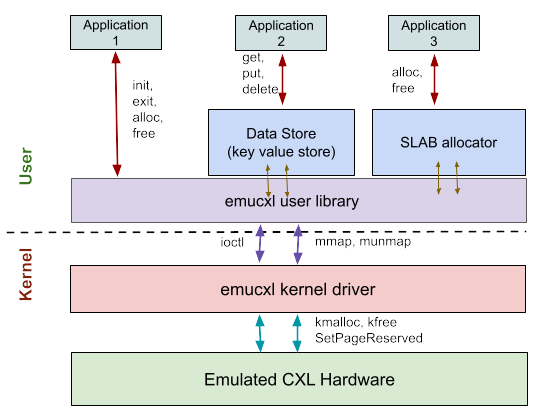}
    \caption{Application use cases using different abstractions 
    implemented via that \Emucxl library.
}
    \label{fig:usecases}
\end{figure}

\subsection{Direct access usage}
\label{subsection:direct}
\begin{table}[tp]
\centering
\ra{1.3}
\begin{tabular}{@{}|c|c|c|c|c|@{}}
% \begin{tabular}{|r|c|c|c|}
\hline
Execution & \multicolumn{2}{c|}{Enqueue} & \multicolumn{2}{c|}{Dequeue} \\
Time (ms) & Local & Remote & Local & Remote\\
\hline
\hline
Mean & 502.98 & 567.21 & 417.69 & 500.40\\
Std. Dev. & 9.23 & 7.93 & 8.71 & 3.66\\
\hline
% \multicolumn{4}{l}{$^{\mathrm{a}}$Sample of a Table footnote.}
\end{tabular}
% \end{center}
\caption{Execution time characteristics of enqueue
and dequeue operations on local and remote memory. 
Reported timings are for 15000 operations each.}

\label{table:direct}
\end{table}

Application 1 shown in Figure~\figref{fig:usecases}
represents an application that operates in 
direct access mode to utilize the \Emucxl interface.

Such applications make calls to the \Emucxl 
library using the user space APIs (shown in Table~\tableref{table:apis}).
Further, the provisioning and management 
of both local and remote memory 
is embedded within the logic of the applications.

To demonstrate the \textit{direct access} use case, 
we implemented a simple queue data structure using the 
\Emucxl library. The queueing application embeds logic
to choose whether objects should be stored in the 
local memory or the remote memory. 
Currently, as part of our simple implementation, 
as part of queue initialization, the application can choose
whether the queue should be maintained entirely local 
or remote memory. This design can be easily extended to
implement more subtle user-space policies that manage
the local and remote memory in an unified manner, via 
promotions and demotions to/from local memory. 

\begin{lstlisting}[language=myc,breaklines=true,
caption={Enqueue and Dequeue functions of a queue with the \Emucxl library.},label={lst:queue}]
struct node* createNode(struct Queue *que, int data)
{
  struct node *newnode;
  newnode = (struct node*)emucxl_alloc(sizeof(struct node),que->policy);
  if(newnode != NULL){
    newnode->data = data;
    newnode->next = NULL;
  }
  return newnode;
}
int enqueue(struct Queue *que, int data)
{
  struct node *newnode = createNode(que, data);
  if((que->front == NULL) && (que->rear == NULL)) 
    que->front = que->rear = newnode;
  else { 
    que->rear->next = newnode;
    que->rear = newnode;
  }
  que->count++;
  return 1;
}
int dequeue(struct Queue *que)
{
  struct node *temp;
  if((que->front == NULL) && (que->rear == NULL))
    return 0;
  temp = que->front;
  que->front = que->front->next;
  if(que->front == NULL)  que->rear = NULL;
  emucxl_free(temp, sizeof(struct node));
  que->count--;
  return 1;
}
\end{lstlisting}
In this use case, the queue is implemented as a singly linked 
list. Each enqueue operation creates a new node using the \textit{emucxl\_alloc} operation, and dequeue and
queue destruction operations involve deleting and freeing 
each node with the help of \textit{emucxl\_free} operation.
Listing~\ref{lst:queue} shows the function definitions
of the enqueue and dequeue operations using the \Emucxl API
calls for using emulated CXL-based disaggregated memory
for managing objects of the queue.

As evaluation of this use case, we conducted an experiment 
with \Emucxl virtual application and measure the execution time
for 15000 queueing operations.
Table~\tableref{table:direct} reports the mean and standard deviation of the total time for the 15000 operations in two setups---all operations
in local memory and all operations in remote memory. 
The results show that operations on the remote memory are marginally costly
and mimic the expected NUMA-like latency characteristics of 
CXL hardware in the memory pooling mode.

\subsection{Middleware driven usage}
\label{subsection:middleware}
In this category of uses cases, we discuss two examples. \\

%%%%%%%%%%%%%%% KEY VALUE STORE %%%%%%%%%%%%%%%

\textbf{Key-value store.} 
Figure~\figref{fig:usecases} shows an application category (Application 2) that
use the key-value middleware for its usage, and the key-value
store in turn interacts with the \Emucxl library 
for managing objects of the store across local and remote memory. 
Key-value stores are a critical component of storage 
infrastructure in a wide range of applications---caches,
metadata stores, collections of items etc~\cite{key-value}.
 Redis\cite{redis} and RocksDB\cite{rocksdb} are open source,
in-memory and persistent key-value stores, respectively.
For both of these widely used stores, CXL can 
increase the memory capacity with low latency and high bandwidth,
via the memory pooling approach.
As part of our use case, a key-value store uses the \Emucxl library
and its API to provide the get, put, delete semantics to applications.\\

\begin{lstlisting}[language=myc,breaklines=true,
caption={key-value store middleware: Pseudo code for PUT function using \Emucxl library. Assume, remote memory is sufficiently large.},label={lst:kvs:PUT}]
void put(kv_store* kvs, char* key, char* value) {
    /* 
    * Create a new key value object(kvs_obj)
    * in the local memory, object stores
    * key-value pair(kvs_pair) and some metadata
    */
    kvs_obj* obj =(kvs_node*)emucxl_alloc(sizeof(kvs_obj), LOCAL_MEMORY);
    obj->kv_pair = (kvs_pair*)emucxl_alloc(sizeof(kvs_pair), LOCAL_MEMORY);
    
    strcpy(obj->kv_pair->key, key);
    strcpy(obj->kv_pair->value, value);
    /*
    * Insert the object at the head (Most recently
    * used position)
    * Update the number of local objects 
    * Evict the object at the tail (Least 
    * Recently Used)  if the size exceeds the limit 
    * and move the evicted object to remote memory
    * update the number of local and remote objects
    */
}
\end{lstlisting}

\begin{lstlisting}[language=myc,breaklines=true,
caption={key-value store middleware: pseudo code for GET function using \Emucxl library. Implementation of policies are not shown here to avoid clustering.},label={lst:kvs:GET}]
const char* get(kv_store* kvs, char* key) {
  /* Search(key_search) in the local objects */
  kvs_node* curr = key_search(kvs->local_head, key);
  if(curr != NULL) { // Found
    return curr->kv_pair->value;
  } else {
    /* Search in the remote objects */
    curr = key_search(head, key);
    if (curr != NULL) { // Found
    /* Implementation of policies to handle GET requests if objects is in remote */
      return curr->kv_pair->value;  
    } else {
      return NULL;
    }
  }
}
\end{lstlisting}

\begin{lstlisting}[language=myc,breaklines=true,
caption={key-value store middleware: pseudo code for DELETE function using \Emucxl library.},label={lst:kvs:DELETE}]
void free_object(kvs_obj* curr)
{
    /* Remove the object from the list of objects,
    * free the object
    * and update the number of objects */
    emucxl_free((void*)curr->kv_pair, sizeof(kvs_pair));
    emucxl_free((void*)curr, sizeof(kvs_obj));
}
void delete(kv_store* kvs, char* key) {
  /* Search(key_search) in the local objects */
  kvs_obj* curr = key_search(kvs->local_head, key);  
  if(curr != NULL) { //Found
    free_object(curr);
  }
  else { /* Search in the remote objects */
    curr = key_search(kvs->remote_head, key);
    if(curr != NULL) { //Found
        free_object(curr);
    }
  }
}
\end{lstlisting}

Listings~\ref{lst:kvs:PUT},~\ref{lst:kvs:GET} and~\ref{lst:kvs:DELETE} 
list function definitions for the PUT, GET, DELETE
operations of the key-value store.
As can be seen, all three functions implement semantics
of the object access and storage for PUT, GET and DELETE
using \Emucxl API calls. The emulation of the CXL-based disaggregated memory is
transparent to the data store operations and metadata handling.

\begin{table}[t]
\centering
\ra{1.3}
\begin{tabular}{@{}c|c|c|c@{}}
% \begin{tabular}{|c|c|c|c|}
\toprule
\textbf{90\% get requests} & \multicolumn{2}{c|}{\% Local fetches} & \\
\textbf{to \% objects} &  Policy 1 & Policy 2 & difference \\
\midrule
10\% & 81.37\% & 3.29\% & 78.08\% \\
\hline
20\% & 50.95\% & 3.77\% & 47.18\% \\
\hline
30\% & 28.59\% & 4.28\% & 24.30\% \\
\hline
40\% & 18.03\% & 4.94\% & 13.09\% \\
\hline
50\% & 14.87\% & 5.94\% & 8.93\% \\
\hline
60\% & 12.67\% & 7.57\% & 5.10\% \\
\hline
70\% & 12.68\% & 10.00\% & 2.68\% \\
\hline
80\% & 22.22\% & 21.17\% & 1.05\% \\
\hline
90\% & 30.43\% & 29.95\% &  0.48\% \\
\hline
Random Access & 29.79\% & 30.01\% & -0.22\% \\
\bottomrule
\end{tabular}
\caption{Comparison two policies to handle 
GET requests of a key-value store. Policy 1 is 
optimistic and moves objects to local memory on access,
Policy 2 is conservative and does not move objects on access.
}
\label{table:key-value-store}
\end{table}

We also demonstrate the flexibility
to implement different policies to handle local and remote
memory for the middleware.
For this purpose, we implemented two simple policies to 
handle GET requests namely Policy1 and Policy2. 
With Policy1, on each GET request, the middleware
first searches for the requested object in the local 
memory and if not found, in the remote memory.
If the object is found in remote memory, the object is 
optimistically moved to the local memory, akin to caching
for subsequent access.
On the other hand, 
Policy2 follows a more conservative approach where the 
middleware simply searches for and retrieves the 
requested object without any additional data movements.

To demonstrate the correctness of our implementation, 
we used a simple setup---local memory capacity of 300 objects
and 1000 objects for remote memory. All PUT operations
added objects to the local store and moved objects (if required)
to the remote store using a Least Recently Used (LRU) policy.

Table~\tableref{table:key-value-store} summarizes the comparison of
both policies with 1000 PUT requests, followed 
by 50000 GET requests to the key-value store. 
As can be seen, if a small set of objects are frequently accessed,
as expected, Policy1 results in a higher number of local memory
fetches. As the trend of access spreads over the entire set of 
objects, the performance of both policies tends to be similar.

This use case presents only a limited and simple feature
set for the \emucxl-based key-value store and is intended to
demonstrate the relevance and applicability of the \Emucxl library. 
Developers can easily customize, adapt, and extend the functionality 
of the key-value store for further involved operations and features. \\

%%%%%%%%%%%%%%%%%%% Slab Allocator %%%%%%%%%%%%%%%%%%%%%%%%%%%%%%

\textbf{Slab allocator.} 
Slab allocation \cite{slaballocator} is a memory management strategy that
improves the allocation of objects in memory. 
The aim is to decrease the impact of fragmentation caused by repetitive allocations and deallocations, and thereby improve
memory utilization. 
The fundamental building block of a slab  allocator is the slab. 
A slab is comprised of one or more virtually contiguous memory pages, 
which are further divided into equal-sized chunks. Each chunk 
represents a unit of memory that can be allocated to hold an object, 
and a reference count is maintained to track the number 
of allocated chunks within the slab.  The use of slabs in 
memory allocation offers several significant advantages---easy
reclamation of unused memory, constant time allocation and deallocation 
and minimal internal fragmentation.
\\
Figure~\figref{fig:usecases} shows another class of applications 
(labelled Application 3) which issue memory allocation and 
deallocation requests via a slab allocator. The user space
allocator can itself be implemented using the \Emucxl library.
For example, the slab allocator can allocate slabs either
from local or remote memory and move around objects 
across different regions for slab management. The slab allocator managing local and remote memory 
ensures that applications benefit from slab optimizations 
without direct involvement of memory provisioning across a memory pool.
While our current implementation does not include the slab 
allocator, we plan it for future release.

Furthermore, our current \Emucxl library implementation, uses
the \verb|mmap| call to map memory into user space at the
granularity of pages. A slab allocator can optimize memory 
usage by allocating page-aligned regions, and allocating
small regions to user level memory requests.

%%%%%%%%%%%%%%%%%%%%%%%%%%%%%%%%%%%%%%%%%%%%%%%%%%%%%%%%%%%%%%%%%%%%%%%%%%%%%%
%%%%%%%%%%%%%%%%%%%%%%%%%%%%%%%% RELATED WORKS %%%%%%%%%%%%%%%%%%%%%%%%%%%%%%%
%%%%%%%%%%%%%%%%%%%%%%%%%%%%%%%%%%%%%%%%%%%%%%%%%%%%%%%%%%%%%%%%%%%%%%%%%%%%%%
\section{Related Work}
\label{section:related_works}

CXL~\cite{cxl} is an open, industry-backed standard which 
enables cache-coherent interconnects for processors, memory and 
accelerators. The CXL consortium~\cite{cxlconsortium} has released
three standards---1.0/1.1~\cite{cxl1.1}, 2.0~\cite{cxl2.0}, and 
3.0~\cite{cxl3.0}. CXL has the benefit of supporting both
standard PCIe devices as well as CXL devices---
all on the same link, enabling backward 
compatibility and leveraging existing infrastructure investments.
Given the prominence of CXL as a potential future 
interconnect technology, many commercial vendors are actively 
engaged in the development of hardware products and prototypes 
for CXL devices, as well as software solutions based on CXL 
technology.
% \teal{Given its popularity promising visions, CXL has attracted 
% much attention in the computing world, 
% many commerical vendors are in the fray
% to develop hardware product/prototypes of cxl 
% devices or cxl-based software solutions.}
In line with this trend, Samsung has reported the development
of prototypes for CXL 1.1 and CXL 2.0 based DRAM for server 
platforms~\cite{samsungcxl1.1, samsungcxl2.0}.
The Leo Smart Memory Controller by Astera Labs also 
reports support for memory expansion, 
memory pooling and memory sharing~\cite{leo-astera}.
The testing and delivery timelines for these products
are expected to be by the end of the year.
Furthermore, Samsung’s CXL Memory Expander hardware is also
accompanied with SMDK~\cite{smdk}, Scalable Memory Development Kit,
which is toolkit of tools and libraries for applications to
use CXL memory. \Emucxl, in spirit, is similar to SMDK, and
can be used with CXL emulation platforms.

While CXL hardware is awaited, work towards software support 
for CXL and and implications on applications has shown lot of interest.
Linux support for CXL devices is in active development~\cite{libcxl, libcxl2}
and already provides kernel support and user interface
tool and libraries to access CXL compliant devices.
Additionally, QEMU supports CXL emulated devices 
for hosts and virtual machines (along with KVM)~\cite{qemucxl, qemucxlsteve}. However, it is 
worth noting that load/store instruction
is not yet supported on the emulated devices. Although 
a lot of the emulation interfaces are in place, integration
with physical or emulated CXL devices is still missing. Unlike these efforts, \Emucxl simultaneously provides both a 
standardized API as an user-level interface for applications 
and a CXL emulation platform. As support and integration with 
physical CXL devices and/or standard CXL emulation platforms
evolves, \Emucxl can be extended and integrated with them,
remaining relevant for rapid prototyping and application usage.

Aside from the software library and tools for CXL,
CXL has the potential to impact a wide ranges of 
applications, middleware and management solutions. A brief list of some of the examples are as follows,
design of disaggregated memory solutions 
and memory pooling~\cite{pond, wahlgren2022,cxl-usecase,directcxl},
cache-coherent integration of accelerators (GPUS, FPGAs) with CPUs~\cite{cxl1.1,cxl2.0, cxl3.0}, 
low-latency and high bandwidth data sharing 
between different types of processors, accelerators 
and memory devices~\cite{dd}, facilitating heterogeneous computing~\cite{cxl-heterogenous}. 
These implications and solutions have the potential
to disrupt existing date center and production infrastructure
to disaggregated and rack-scale computing setups. \Emucxl complements these efforts by providing a standardized framework to rapidly develop CXL-based solutions. Furthermore, the emulation setup
and library can itself be extended to accommodate a broader range of CXL semantics and operations. We believe that \Emucxl
can fulfills an important need for developers
and researchers, and to aid exploration of CXL's
implications and benefits.

%%%%%%%%%%%%%%%%%%%%%%%%%%%%%%%%%%%%%%%%%%%%%%%%%%%%%%%%%%%%%%%%%%%%%%%%%%%%%%
%%%%%%%%%%%%%%%%%%%%%%%%%%%%%%%% CONCLUSION %%%%%%%%%%%%%%%%%%%%%%%%%%%%%%%%%%
%%%%%%%%%%%%%%%%%%%%%%%%%%%%%%%%%%%%%%%%%%%%%%%%%%%%%%%%%%%%%%%%%%%%%%%%%%%%%%

\section{Conclusion \textit{\&} Future Work}
\label{section:conclusion}

As a once-in-a-decade technological force, the rapid adoption of CXL
is expected to provide significant benefits for various use cases--
from disaggregated memory applications, such as large-scale in-memory 
databases, to real-time analytics and computationally
intensive applications.
As part of this work, we addressed the need for rapid prototyping
and a platform for CXL based solutions in a disaggregated memory 
setup. We developed \Emucxl that ships as a
open source library and the 
standardized user space API for use with an emulated CXL platform along with virtual machine setup.
We discussed the implementation details of the \Emucxl library,
demonstrated its relevance and features via two types of 
implementation use cases of disaggregated memory applications
and their evaluation. Our solution is open sourced and can
be easily used to extend its feature set both from the library
perspective and also build new middleware using the library.
We believe that the \Emucxl fills an important gap to 
implement new CXL emulation and library capabilities and
also for quick development and testing of CXL based applications.

As part of immediate future work, we plan to add further to
the set of use cases to show case \emucxl's applicability,
e.g., the slab allocator, integration with open-source 
object stores etc. Further, currently, \Emucxl is designed
to work with a single process and needs further management
when multiple entities access and use a shared disaggregated
memory pool. We plan to add features and support for management
operations across multiple processes and disaggregated memory
via the \Emucxl library.

% %%%%%%%%%%%%%%%%%%%%%%%%%%%%%%%%%%%%%%%%%%%%%%%%%%%%%%%%%%%%%%%%%%%%%%%%%%%%%%
% %%%%%%%%%%%%%%%%%%%%%%%%%%%%%%%% Acknowledgements %%%%%%%%%%%%%%%%%%%%%%%%%%%%%%%%%%
% %%%%%%%%%%%%%%%%%%%%%%%%%%%%%%%%%%%%%%%%%%%%%%%%%%%%%%%%%%%%%%%%%%%%%%%%%%%%%%

% \section*{Acknowledgements}
% \label{section:acknowledgements}
% The authors would like to thank the anonymous
% reviewers for their insightful comments and suggestions.

% \puru{for paper references from workshops, conferences and journals, do not need to mention the URL.}

% \raja{should i remove the url then?}

\bibliographystyle{ieeetransN}

\end{document}